# Giant and robust topological Hall effect in Chiral Magnet $Co_7Zn_8Mn_5$

Hai Zeng[1,2,3,a], Xuanwei Zhao[1,2,3,4], Guang Yu[1,2,3,a], Xiaohua Luo[1,2], Shengcan Ma[1,2,*], Changcai Chen[1,2], Zhaojun Mo[7], Yugang Zhang[5,6], Yisheng Chai[5,6,* *], Jun Shen[7,8,9], and Zhenchen Zhong[1,2]

[1] *Jiangxi Key Laboratory for Rare Earth Magnetic Materials and Devices, College of Rare earths, Jiangxi University of Science and Technology, Ganzhou 341000, PR China.*

[2] *Ganzhou Key Laboratory for Rare Earth Magnetic Functional Materials and Physics, College of Rare earths, Jiangxi University of Science and Technology, Ganzhou 341000, PR China.*

[3] *School of Materials Science and Engineering, Faculty of Materials Metallurgy and Chemistry, Jiangxi University of Science and Technology, Ganzhou 341000, PR China.*

[4] *China Key Laboratory of Radiation Physics and Technology and Key Laboratory of High Energy Density Physics and Technology of Ministry of Education, Sichuan University, Chengdu, 610064, PR China.*

[5] *Low Temperature Physics Laboratory, College of Physics, Chongqing University, Chongqing, 401331, PR China.*

[6] *Center of Quantum Materials and Devices, Chongqing University, Chongqing 401331, PR China.*

[7] *Ganjiang Innovation Academy, Chinese Academy of Sciences, Ganzhou, Jiangxi 341000, PR China.*

[8] *Key Laboratory of Cryogenics, Technical Institute of Physics and Chemistry, Chinese Academy of Sciences, Beijing 100190, PR China.*

[9] *University of Chinese Academy of Sciences, Beijing 100049, PR China.*

**Abstract**   Recently, *β*-Mn-type Co-Zn-Mn alloys have gained particular attentions as a new class of chiral magnets hosting skyrmion phase. In this work, a giant topological Hall effect (THE) is observed during the wide temperature range below 220 K in the chiral magnet $Co_7Zn_8Mn_5$. The maximum topological Hall resistivity, -2.1 μΩ cm, is obtained at 10 K. Moreover, the observed THE effect persists up to $T_c$, which is mainly derived from the noncoplanar spin structure with scalar spin chirality. In contrast, the formation of skyrmion phase is substantiated at the temperature interval slightly below $T_c$ by adopting the magnetization and ac-susceptibility. Further, the possible signal of skyrmion-conical coexisting phase is found based on the out-of-phase component in magnetoelastic measurements. These results strongly suggest the chiral magnet $Co_7Zn_8Mn_5$ compound should be an excellent candidate to study the topological magnetic properties and high temperature skyrmions.

**Corresponding author:**

*Authors to whom correspondence should be addressed:

shengcanma801@gmail.com, mashengcan@jxust.edu.cn (Shengcan Ma) and yschai@cqu.edu.cn (Yisheng Chai).

[a)] These authors contributed equally: Hai Zeng, Guang Yu

# 1. Introduction

Topologically protected nontrivial magnetic textures have recently become one of the frontiers and hot spots in condensed matter physics [1,2], showing promising application prospect for spintronic devices, such as unconventional computing devices [3] and racetrack memory devices [4]. In these magnetic systems, topological Hall effect (THE) can usually be observed in real space [5,6]. The origin of the THE can be characterized by a non-zero scalar spin chirality, $\chi_{ijk} = \mathbf{S}_i \cdot (\mathbf{S}_j \times \mathbf{S}_k)$. Here, $\mathbf{S}_i$, $\mathbf{S}_j$ and $\mathbf{S}_k$ represent the nearest localized spin moments respectively [6-8]. A quantum-mechanical Berrry phase, whose phase factor is proportional to the $\chi_{ijk}$, collected by conduction elections when these elections pass through the noncoplanar spin configurations. The Berry phase acts as the fictitious magnetic field and induces an additional contribution to Hall resistivity as termed THE [7-10].

Actually, the THE induced by spin chirality is usually counteracted by the sum over the whole lattice sites. There are two mechanisms to avoid the THE being offset in real materials [11]. Magnetic materials, crystallizing in nontrivial crystal structure, certainly have inequivalent multiple loops in a unit cell [6]. Hence, a Berry phase is collected by the electrons, which reflects the winding number of each band [11,12]. The additional Hall conductivity $\sigma_{xy}^T \sim \sum f(e) \boldsymbol{B}_k^{em}$ is the sum of all band states, where $\boldsymbol{B}_k^{em}$ and $f(e)$ are the fictitious magnetic field and Fermi distribution function [13]. The THE induced by this mechanism has been observed in many materials with pyrochlores structure, such as $Nd_2Mo_2O_7$ [13] and $Pr_2Ir_2O_7$ [14]. Spin-orbit interaction for ferromagnetic materials also gives rise to net Berry curvature in momentum space, which induces anomalous Hall effects (AHE) [15]. Another mechanism is associated with nonzero topological spin texture, *e.g.*, skyrmion (Sk) phase. This phenomenon is usually observed when the modulation period of spin texture is much larger than the lattice constants of compounds [16]. In this case, the topological winding induces a real-space quantum Berry curvature (acting as fictitious magnetic field $\boldsymbol{B}_r^{em}$) for the conduction electrons. The magnetic field $B_r^{em}$ would lead to the Lorentz force $F = -e\boldsymbol{v} \times \boldsymbol{B}_r^{em}$ and accordingly Hall resistance $\rho_{xy}^T \sim \boldsymbol{B}_r^{em}$ [8-10]. Therefore, the THE is one of the significant electrical transport signals exhibited by the Sks [5,8].

The first experimental observation of magnetic Sk phases has been reported in chiral B20 crystals by means of small-angle neutron scattering and Lorentz transmission electron microscopy, such as MnSi and FeGe [17,18]. In B20 compounds, the interplay between Dzyaloshinskii-Moriya interactions (DMI) arising from the strong spin-orbit coupling, Heisenberg exchange interaction, and magnetic anisotropy can stabilize the Sk lattice under external magnetic fields at the narrow *T-H* window just below magnetic-ordering temperature $T_c$ [19,20]. But a defect of B20 crystals is their very low Curie temperature, which precludes their application at room temperature. Ternary alloys Co-Zn-Mn alloys are DMI-based Sks-hosting materials, which is a typical material system to resolve this conundrum for their large alteration in Curie temperature $T_c \approx$ 148-462 K by controlling Mn content [21,22]. The exploration and study of magnetic properties associated with skyrmionic behavior have been a main research focus on the Co-Zn-Mn system during the past decade [23-27]. However, few investigations about novel transport properties in Co-Zn-Mn compounds, such as THE, have not been reported thus far. In this paper, the magnetic and transport properties are investigated in $Co_7Zn_8Mn_5$ alloy. We report a large THE in a broader temperature range of 10-220 K in $Co_7Zn_8Mn_5$ alloy, demonstrating its robustness, which should be caused by the complex spin configurations, not just Sk textures.

## 2. Experimental methods

The polycrystalline samples of $Co_7Zn_8Mn_5$ were synthesized by an improved sintering method.

The stoichiometric amounts mixtures of Mn pieces (purity 99.98%), Co and Zn pellets (purity 99.99%) were sealed in quartz tube under an argon atmosphere. The quartz tube was put into a furnace, quickly heated up to 1273 K where it remains during 24 h. Then, the sample was slowly cooled (1 K/h) downed to 1198 K and kept for 100 h at this temperature, after which the quartz tube was quenched into water. Samples for transport measurements were cut into rectangular sheet with smooth surface and the sample aspect ratio is 3:1 to 5:1 to ensure a uniform electric field during the electrical measurements. The crystal structure and phase purity were identified by using the X-ray powder diffraction (XRD) with Cu-$K_\alpha$ radiation ($\lambda$ = 1.54184 Å). The XRD pattern was analyzed by Rietveld refinement [28]. The magnetic and electrical transport properties were measured by using the VSM (vibrating sample magnetometer) and ETO (electron transport option) module, respectively, equipped in a DynaCool Cryogen-free Physical Property Measurement System (PPMS, Quantum Design, USA). The ac susceptibility $\chi = \chi'+i\chi''$ were measured by using the Magnetic Property Measurement System (MPMS-XL Quantum Design SQUID magnetometer). To eliminate the experimental errors, the raw data of longitudinal ($\rho_{xx}$) and transverse ($\rho_{yx}$) resistivities were evaluated using the following formulas [10]:

$$\rho_{xx}(\mu_0 H) = \left[\rho_{xx}(+\mu_0 H) + \rho_{xx}(-\mu_0 H)\right]/2 \quad (1)$$

$$\rho_{yx}(\mu_0 H) = \left[\rho_{yx}(+\mu_0 H) - \rho_{yx}(-\mu_0 H)\right]/2 \quad (2)$$

It has been reported that magnetostriction behavior in MnSi single crystal would be related to the existence of the Sk phase under a lower applied field [29]. Thus, ac magnetoelectric (ME) response should also be taken as evidences of the Sks formation. ac ME response of Sk phase was detected in a composite magnetoelectric configuration consisting of $Co_7Zn_8Mn_5$ alloy and thin piezoelectric PMN-PT ($0.7Pb(Mg_{1/3}Nb_{2/3})O_3$-$0.3PbTiO_3$) that was bonded together using the silver epoxy, based on piezoelectric transducer technique [30]. The driven frequency is 1 kHz. The ME responses can be described by the following formula:

$$\Delta E = \alpha_E \Delta H \tag{3}$$

in which $E$, $\alpha_E$ and $H$ are the electric field, ME coefficient and magnetic field, respectively. Notably, the $\alpha_E$ with out-of-plane electric field and magnetic field ($\alpha_{EO}$) is more sensitive when detecting different Sk phases, including the skyrmions-conical coexisting phase induced by low density or static disorder [31].

## 3. Results and discussion

### 3.1 Crystal structure of $Co_7Zn_8Mn_5$

The Co-Zn-Mn system crystallizes in the $\beta$-Mn-type crystal structure, as shown in Fig. 1a. The space group of this structure is $P4_132$ (the left side of Fig. 1a) or $P4_332$ (the right side of Fig. 1a), depending on its chirality. There are 20 atoms in one unit cell, which are classified into two kinds of crystallographic sites, $8c$ and $12d$ Wyckoff positions, which are occupied by Co atoms and other atoms, respectively [22,32]. Figure 1b shows the Rietveld refinement of the XRD pattern for $Co_7Zn_8Mn_5$ alloy. Crystal structure of $\beta$-Mn-type and impurity free phase and were confirmed. The lattice parameter $a$ was calculated to be 6.391 Å, being larger than that of Mn-free $Co_{10}Zn_{10}$ alloy ($a$ = 6.323 Å) [22]. The real chemical composition was determined to be $Co_{7.3}Zn_{8.0}Mn_{4.58}$ by using the energy dispersive X-ray spectroscopy (Fig. S1), which is very close to its nominal composition.

### 3.2 Magnetic properties and evolution of the spin textures

Shown in Fig. 2a are the temperature dependent magnetization $M(T)$ curves upon zero-field-cooled (ZFC) and field-cooled (FC) processes for $Co_7Zn_8Mn_5$ compound under an applied field of $H$ = 200 Oe. During the FC process, the magnetization exhibits a rapid increase at the magnetic transition temperature $T_c \sim$ 220 K, and then gradually decreases from 150 K to 28 K owing to the increase of helical $q$ [33]. As partial substitution of Mn proceeds to Co, the $T_c$ falls quickly ($T_c$ of $Co_{10}Zn_{10}$ is ~ 415 K [21] or 470 K [22]). On the other hand, the magnetization rises sharply around $T_g \sim$ 20 K during the ZFC branch due to the reentrant spin-glass transition [34]. The spin glass states indicate magnetic frustration effects of $Co_7Zn_8Mn_5$ alloy inherited from $β$-Mn [21].

The isothermal magnetization loops are measured and presented in Fig. S2. It is found that the magnetization reaches saturation very rapidly below $T_c$ with low saturating field. Furthermore, it is of significance that the magnetization behavior without a hysteresis is observed in the whole temperature range covered in this work. To probe the signal of the Sks formation in $Co_7Zn_8Mn_5$ compound, the magnetic field dependence of isothermal d$M$/d$H$ curves were measured at different temperatures near $T_c$ (Fig. 2b). A well-defined dip/peak can be discovered under magnetic field of 0.022 T in temperature range from 212 K to 216 K, which is a hallmark associated with the formation and annihilation of Sk phase with the evolution of magnetic field [21]. As expected, the Sks start to appear just below the magnetic-ordering temperature, a narrow temperature interval of about 4 K, suggesting that thermal fluctuations play a crucial role in the stablizaton of Sk phase under competing interactions with similar energy (*e.g.* DMI, Heisenberg exchange interaction and magnetic anisotropy) [17].

To further confirm the formation of the Sks in $Co_7Zn_8Mn_5$ compound, we have performed ac susceptibility measurements. Displayed in Fig. 3a are the real part of magnetic susceptibility $\chi'$ as a function of external magnetic field at selected temperatures just below $T_c$ after the ZFC process. A similar dip/peak anomaly induced by the magnetic field in the $\chi'(H)$ should be discerned inside the temperature range between 211 K and 218 K and fades out outside the temperature region, which is found in other chiral Co-Zn-Mn alloys [21]. In fact, the dip anomaly is more straightforward to justify the formation of Sk phase in the imaginary part $\chi''(H)$ (shown in Figs. 3d and S3). More detailed data of ac susceptibility are shown in Fig. S3. Here, the evolution of the magnetic structure with the magnetic field for the sample can be distinguished clearly in Figs. 3b, c and d. Let's take $T$ = 216 K as an example, the d$M$/d$H$, $\chi'$ and $\chi''$ *versus* applied field are given in Fig. 3. It should be noted that the pronounced anomaly appears between the critical fields $H_{a1}$ (~0.01 T) and $H_{a2}$ (~0.02 T), for field dependence of d$M$/d$H$, $\chi'$ and $\chi''$ plots. As seen in Fig. 3(d), an initial increase of $\chi''$ induced by external field is related to the transitions from the helical (H) to conical (C) state [35]. With further increasing magnetic field, the magnetic order first enters the Sk phase regime between $H_{a1}$ and $H_{a2}$ [36], then reenters the conical phase, and finally into the polarized ferromagnetic (FM) state. In other words, four transitions should occur in sequence with the evolution of magnetic field: H→C→Sk→C→FM. Such a similar behavior has been also reported in chiral crystals such as FeGe [35], $Fe_{0.7}Co_{0.3}Si$ [36] and $Cu_2OSeO$ [37], albeit the extrema of $\chi'$ and $\chi''$ do not correspond exactly with the inflection points of d$M$/d$H$ at the boundary of the Sk phase.

*3.3 Magnetoelectric properties and skyrmions signature*

Figure 4a and 4b reveal the magnetic field dependence of ac ME coefficients $α_x$ and $α_y$ at different temperatures just below $T_c$. With increasing magnetic field, $α_x$ shows a smooth downward trend below ~0.0634 T at 211 K, while $α_y$ is almost field independent. This situation corresponds to the H ground state. And then, a dramatic enhancement of $α_x$ with further increase of field occurs due to the transition from C state to field-induced FM order. When the temperature increases to 213.5 K, a low field peak/dip anomaly emerges in both $α_x$ and $α_y$ *versus* external field curves, which may be related to the appearance of Sk phase. It is worth noting that the peak/dip features remain until to ~217 K. The *T-H* window of the Sk phase agrees well with the result of ac susceptibility. Above 217

K, nevertheless, it's not clear about the magnetic structure for the weak signal of ac magnetoelastic response.

To probe the field evolution of the Sk phase, the Argand diagrams derived from Figs. 4a and b are shown in Fig. 5. It is observed that the Argand curves away from Sk phase show a nearly zero phase angle, validating the reliability of our test results. At 217 K (below but near $T_c$), it is noted that the curve shows a straight line under the ac driven field, implying a no dissipative behavior for C or FM phase. In fact, only the Sk phase can show the non-dissipative feature in $Co_7Zn_8Mn_5$ alloy. Below 216.5 K, semicircle features anomaly as the electric signal of out-of-phase component start to appear and evolve into circle features. The magnetic field range of dissipating behavior agrees well with the peak anomaly in the imaginary part of ac susceptibility. At lower temperature region, the circle features gradually shrink and submerge in the baseline. Similar dissipating behavior can be observed in MnSi single crystal [31], but there is no evidence for the existence of pure skyrmion lattice. The existence of a large amount of grain boundaries and defects maybe destroy the ordered skyrmion lattice in the polycrystalline $Co_7Zn_8Mn_5$ sample. Therefore, it is concluded that the out-of-phase component should be associated with the coexistence of Sk and C phases (Sk-C).

Figure 4c and 4d exhibit the ac ME coefficient $α_x$ and $α_y$ curves as a function of temperature, which are measured under different magnetic field. The ME coefficient $α_x$ and $α_y$ exhibits the opposite trend under different magnetic field. Clear peaks of $α_x$ (corresponding to the dips in $α_y(T)$ curves) are observed ~220 K, marking the progress of magnetic ordering transition. In combination with Figs. 4 a and b, applying the magnetic field induces a dip or peak in the corresponding position of $α_x(T)$ and $α_y(T)$ curves denoted by the dotted line. It is worth noting that this peak in Fig. 4d shifts to lower temperature and becomes more noticeable upon further increasing the field. Very evidently, the peak feature points to the onset of Sk-C coexistence phase just below $T_c$, where fluctuations become predominant [36]. Note that the characteristic peak disappears at 0.03 T, which corresponds well with the results of ac susceptibility as well.

*3.4 Topological Hall effect of $Co_7Zn_8Mn_5$*

So far, the Sk phase has been observed in several Co-Zn-Mn compounds during the past years. But we notice that there are little investigations about the transport properties, especially the THE, to be reported in Co-Zn-Mn systems yet. As discussed above, the formation of Sks is anticipated in the $Co_7Zn_8Mn_5$ alloy. Furthermore, large THE has been regarded as the hallmark of formation of Sk phase [6,8,10]. Here, the transport measurements were carried out on the $Co_7Zn_8Mn_5$ compound to reveal the possible contribution from THE effects. Figure 6a shows the temperature dependence of longitudinal resistivity curves upon heating under zero magnetic field. The resistivity takes on a typical metallic behavior with a large residual resistance, which is consistent with the results reported in the literature [21]. At low temperature regime, it should be noted that the resistance reduces slightly with increasing temperature up to $T_g$, manifesting the disorder features of magnetic moment below $T_g$, which agrees well with the ZFC $M(T)$ measurements (Fig. 2a). With further enhancing the temperature, the resistivity value increases monotonously following by a clear kink point at $T_c$, demonstrating the transition to the PM state. Figure 6b shows the magnetic field dependent magnetoresistance {$MR = [\rho_{xx}(\mu_0H) - \rho_{xx}(0)] / \rho_{xx}(0)$}. The data shows a small negative $MR$ ( < 1%) at the temperature interval of 10-300 K, which is due to the suppression of conduction electron scattering by the application of the magnetic field [38].

Shown in Fig. 6c are the magnetic field dependence of Hall resistivity $\rho_{yx}(H)$ curves at several selected temperatures from 10 to 300 K. For all these temperatures, the $\rho_{yx}$ values increase with increasing field and are rapidly saturated at a low critical magnetic field. It is abnormal that, however, the saturation value of $\rho_{yx}$ decreases with enhancing temperature. For a magnetic material, the contributions of $\rho_{yx}$ can be divided into three parts [39]:

$$\rho_{yx} = \rho_{yx}^{N} + \rho_{yx}^{A} + \rho_{yx}^{T} = R_0\mu_0 H + S_A \rho_{xx}^2 M + \rho_{yx}^{T} \tag{4}$$

where $\rho_{yx}^{N}$, $\rho_{yx}^{A}$ and $\rho_{yx}^{T}$ represent the normal, abnormal and topological Hall resistivity, respectively. $R_0$ is the ordinary Hall coefficient, $S_A \rho_{xx}^2$ is the anomalous Hall coefficient and $M$ is the corresponding magnetization. The value of $\rho_{yx}^{T}$ returns to zero at high magnetic field due to the FM state. Only the contributions of $\rho_{yx}^{N}$ and $\rho_{yx}^{A}$ to the total $\rho_{yx}$ value reserve in this case. The coefficients of $R_0$ and $S_A$ can be extracted from the intercept and the slope of the linear fitting of $\rho_{yx}/\mu_0 H$ versus $\rho_{xx}^2 M/\mu_0 H$ at high-field region. Hence, the $\rho_{yx}^{T}$ can be obtained by deducting both $\rho_{yx}^{N}$ and $\rho_{yx}^{A}$ from $\rho_{yx}$ in low field regions. The extracted $\rho_{yx}^{T}$ at different temperatures are shown in Fig. 6d. The large $\rho_{yx}^{T}$ values are obtained below 220 K. The robustness of this large THE is exhibited with a broader temperature window 10 K-220 K in contrast with MnSi [5]. The $\rho_{yx}^{T}$ in this work seems to be caused by the noncoplanar spin texture with scalar spin chirality, not just including the Sks. Both the absolute value of the maximum topological Hall resistivity $\rho_{max}^{T}$ and saturation field $\mu_0 H_z$ of $\rho_{yx}^{T}$ increase with reducing temperature below 210 K, as shown in the inset of Fig. 6d. At 10 K, the $\rho_{max}^{T}$ reaches up to -2.1 μΩ cm, which is 13 times as large as that for B20 compounds MnGe (≈-0.16 μΩ cm) [6]. In addition, it is obvious that the $\rho_{max}^{T}$ almost reaches saturation below 100 K. The temperature independence of $\rho_{yx}^{T}$ is the characteristic for the topological spin texture, which indicates the giant THE at low temperature regime is probably related to the nonzero scalar spin chirality [6].

## 4. Conclusions

In summary, we have reported the magnetic, electrical and elastic properties of the polycrystalline chiral $Co_7Zn_8Mn_5$ magnet. Magnetization and ac-susceptibility measurements unambiguously reveal the formation of Sk phase under magnetic field below $T_c$. Moreover, the possible signal of Sk-C coexisting phase is found based on the out-of-phase component in magnetoelastic measurements. In contrast, we achieved a giant THE in a wide temperature region with a maximum value of -2.1 $\mu\Omega$ cm at 10 K in $Co_7Zn_8Mn_5$ compound, which may be dominantly derived from the spin chirality from the field induced topological magnetic structure at low temperatures.

**Credit author statement**

H. Zeng and G. Yu conceived and designed the project; X. Zhao and Y. Chai performed the experiments and explained the results; Z. Zhong designed the samples; Z. Mo and J. Shen synthesized the samples; X. Luo, C. Chen and S. Ma performed the result analysis; H. Zeng wrote the manuscript after discussion with S. Ma; S. Ma revised the paper. All authors have approved the final version of the manuscript.

**Acknowledgements**

This work has been financially supported by the National Natural Science Foundation of China (Grant Nos. 52061014, 11974065), the Key Project of Natural Science Foundation of Jiangxi Province (Grant No. 20192ACB20004), the Leading Talents Program of Jiangxi Provincial Major Discipline Academic and Technical Leaders Training Program (Grant No. 20204BCJ22004), the National Natural Science Funds for Distinguished Young Scholar(51925605), the Major Science and Technology Research and Development Special Funds Project of Jiangxi Province (Grant No. 20194ABC28005) and the Open Project awarded by Fujian Provincial Key Laboratory of Quantum Manipulation and New Energy Materials (Grant No. QMNEM2002). Y. C would like to thank Miss G. W. Wang at Analytical and Testing Center of Chongqing University for her assistance.

**Data availability statements**

The data that support the findings of this work are available from the corresponding author upon reasonable request.

**Supplementary information**

The online version contains supplementary material available.

**References**

[1] S. Mühlbauer, B. Binz, F. Jonietz, C. Pfleiderer, A. Rosch, A. Neubauer, R. Georgii, P. Böni, Skyrmion lattice in a chiral magnet, Science 323 (2009) 915.

[2] X. Z. Yu, N. Kanazawa, W. Z. Zhang, T. Nagai, T. Hara, K. Kimoto, Y. Matsui, Skyrmion flow near room temperature in an ultralow current density, Nat. Commun. 3 (2012) 988.

[3] S. Li, W. Kang, X. Zhang, T. X. Nie, Y. Zhou, K. L. Wang, W. S. Zhao, Magnetic skyrmions for unconventional computing, Mater. Horiz. 8 (2021) 854.

[4] S. Parkin, S. H. Yang, Memory on the racetrack, Nat. Nanotechnol. 10 (2015) 195.

[5] A. Neubauer, C. Pfleiderer, B. Binz, A. Rosch, R. Ritz, P. G. Niklowitz, P. Bo¨ni, Topological Hall effect in the A phase of MnSi, Phys. Rev. Lett. 102 (2009) 186602.

[6] N. Kanazawa, Y. Onose, T. Arima, D. Okuyama, K. Ohoyama, S. Wakimoto, K. Kakurai, S. Ishiwata, Y. Tokura, Large topological Hall effect in a short-period helimagnet MnGe, Phys. Rev.

Lett. 106 (2011) 156603.

[7] B. Giri, A. I. Mallick, C. Singh, P. V. P. Madduri, F. Damay, A. Alam, A. K. Nayak, Robust topological Hall effect driven by tunable noncoplanar magnetic state in Mn-Pt-In inverse tetragonal Heusler alloys, Phys. Rev. B 102 (2020) 014449.

[8] X. M. Zheng, X. W. Zhao, J. Qi, X. H. Luo, S. C. Ma, C. C. Chen, H. Zeng, G. Yu, Giant topological Hall effect around room temperature in noncollinear ferromagnet $NdMn_2Ge_2$ single crystal, Appl. Phys. Lett. 118 (2021) 072402.

[9] P. Bruno, V. K. Dugaev, M. Taillefumier, Topological Hall effect and Berry phase in magnetic nanostructures, Phys. Rev. Lett. 93 (2004) 096806.

[10] H. Li, B. Ding, J. Chen, Z. Li, E. Liu, X. K. Xi, G. H. Wu, W. H. Wang, Large anisotropic topological Hall effect in a hexagonal non-collinear magnet $Fe_5Sn_3$, Appl. Phys. Lett. 116 (2020) 182405.

[11] M. Onoda, G. Tatara, N. Nagaosa, Anomalous Hall effect and skyrmion number in real and momentum spaces. J. Phys. Soc. Jpn. 73 (2004) 2624.

[12] K. Ohgushi, S. Murakami, N. Nagaosa, Spin anisotropy and quantum Hall effect in the kagomé lattice: Chiral spin state based on a ferromagnet. Phys. Rev. B 62 (2000) R6065.

[13] Y. Taguchi, Y. Oohara, H. Yoshizawa, N. Nagaosa, Y. Tokura, Spin chirality, Berry phase, and anomalous Hall effect in a frustrated ferromagnet, Science 291 (2001) 2573.

[14] Y. Machida, S. Nakatsuji, Y. Maeno, T. Tayama, T. Sakakibara, S. Onoda, Unconventional Anomalous Hall Effect Enhanced by a Noncoplanar Spin Texture in the Frustrated Kondo Lattice $Pr_2Ir_2O_7$, Phys. Rev. Lett. 98 (2007) 057203.

[15] H. Li, B. Zhang, J. Liang, B. Ding, J. Chen, J. Shen, Z. Li, Large anomalous Hall effect in a hexagonal ferromagnetic $Fe_5Sn_3$ single crystal, Phys. Rev. B 101 (2020) 140409(R).

[16] K. Kadowaki, K. Okuda, M. Date, Magnetization and magnetoresistance of MnSi I, J. Phys. Soc. Jpn. 51 (1982) 2433.

[17] T. Schulz, R. Ritz, A. Bauer, M. Halder, M. Wagner, C. Franz, C. Pfleiderer, K. Everschor, M. Garst, Emergent electrodynamics of skyrmions in a chiral magnet, Nat. Phys. 8 (2012) 301.

[18] X. Z. Yu, N. Kanazawa, Y. Onose, K. Kimoto, W. Z. Zhang, S. Ishiwata, Y. Matsui, Y. Tokura, Near room-temperature formation of a skyrmion crystal in thin-films of the helimagnet FeGe, Nat. Mater. 10 (2011) 106.

[19] N. Kanazawa, S. Seki, Y. Tokura, Noncentrosymmetric magnets hosting magnetic skyrmions, Adv. Mater. 29 (2017) 1603227.

[20] O. I. Utesov, A. V. Syromyatnikov, Cubic B20 helimagnets with quenched disorder in magnetic field. Phys. Rev. B 99 (2019) 134412.

[21] Y. Tokunaga, X. Z. Yu, J. S. White, H. M. Rønnow, D. Morikawa, Y. Taguchi, Y. Tokura, A new class of chiral materials hosting magnetic skyrmions beyond room temperature, Nat. Commun. 6 (2015) 7638.

[22] J. D. Bocarsly, C. Heikes, C. M. Brown, S. D. Wilson, R. Seshadri, Deciphering structural and magnetic disorder in the chiral skyrmion host materials $Co_xZn_yMn_z$ ($x + y + z = 20$), Phys. Rev. Mater. 3 (2019) 014402.

[23] K. Karube, J. S. White, N. Reynolds, J. L. Gavilano, H. Oike, A. Kikkawa, F. Kagawa, Robust metastable skyrmions and their triangular–square lattice structural transition in a high-temperature chiral magnet, Nat. Commun. 15 (2016) 1237.

[24] D. Morikawa, X. Yu, K. Karube, Y. Tokunaga, Y. Taguchi, T. Arima, Y. Tokura, Deformation of topologically-protected supercooled skyrmions in a thin plate of chiral magnet CoZnMn, Nano Lett. 17 3 (2017) 1637.

[25] K. Karube, J. S. White, D. Morikawa, C. D. Dewhurst, R. Cubitt, Disordered skyrmion phase stabilized by magnetic frustration in a chiral magnet, Sci. Adv. 4 (2018) eaar7043.

[26] T. H. Kim, H. Zhao, B. Xu, B. A. Jensen, A. H. King, M. J. Kramer, C. Nan, L. Ke, L. Zhou, Mechanisms of skyrmion and skyrmion crystal formation from the conical phase, Nano Lett. 20 7 (2020) 4731.

[27] V. Ukleev, K. Karube, P. M. Derlet, C. N. Wang, H. Luetkens, D. Morikawa, A. Kikkawa, Frustration-driven magnetic fluctuations as the origin of the low-temperature skyrmion phase in $Co_7Zn_7Mn_6$, npj Quantum Mater. 6 (2021) 40.

[28] H. M. Rietveld, A profile refinement method for nuclear and magnetic structures, J. Appl. Cryst. 2 (1969) 65.

[29] A.E. Petrova, S. M. Stishov, Thermal expansion and magneto-volume studies of the itinerant helical magnet MnSi, Phys. Rev. B 94 (2016) 020410(R).

[30] B. K. Chougule, Magnetoelectric composites, J Mater Sci Technol. 40 (2010) 153.

[31] Y. S. Chai, P. Lu, H. Du, J. Shen, Y. Ma, Probe of skyrmion phases and dynamics in MnSi via the magnetoelectric effect in a composite configuration, Phys. Rev. B 104 (2021) L100413.

[32] W. Xie, S. Thimmaiah, J. Lamsal, J. Liu, T. W. Heitmann, D. Quirinale, $\beta$-Mn-type $Co_{8+x}Zn_{12-x}$ as a defect cubic laves phase: site preferences, magnetism, and electronic structure, Inorg. Chem. 52 (2013) 9399.

[33] K. Karube, K. Shibata, J. S. White, T. Koretsune, X. Z. Yu, Y. Tokunaga, H. M. Rønnow, R. Arita, T. Arima, Y. Tokura, Y. Taguchi, Controlling the helicity of magnetic skyrmions in a $\beta$-Mn-type high-temperature chiral magnet, Phys. Rev. B 98 (2018) 155120.

[34] T. Nakajima, K. Karube, Y. Ishikawa, M. Yonemura, N. Reynolds, J. S. White, H. M. Rønnow, A.

Kikkawa, Y. Tokunaga, Y. Taguchi, Y. Tokura, T. Arima, Correlation between site occupancies and spin-glass transition in skyrmion host $Co_{10-x/2}Zn_{10-x/2}Mn_x$, Phys. Rev. B 100 (2019) 064407.

[35] H. Wilhelm, M. Baenitz, M. Schmidt, Precursor phenomena at the magnetic ordering of the cubic helimagnet FeGe. Phys. Rev. Lett. 107 (2011) 127203.

[36] L.J. Bannenberg, A.J.E. Lefering, K. Kakurai, Y. Onose, Y. Endoh, Y. Tokura, C. Pappas, Magnetic relaxation phenomena in the chiral magnet $Fe_{1-x}Co_xSi$: An ac susceptibility study, Phys. Rev. B 94 (2016) 134433.

[37] F. Qian, H. Wilhelm, A. Aqeel, T. T. M. Palstra, A. J. E. Lefering, E. H. Bruck, C. Pappas, Phase diagram and magnetic relaxation phenomena in $Cu_2OSeO_3$. Phys. Rev. B 94 (2016) 064418.

[38] B. Raquet, M. Viret, E. Sondergard, O. Cespedes, R. Mamy, Electron-magnon scattering and magnetic resistivity in 3d ferromagnets. Phys. Rev. B 66 (2002) 024433.

[39] N. Nagaosa, J. Sinova, S. Onoda, A. H. MacDonald, N. P. Ong, Anomalous Hall effect. Rev. Mod. Phys. 82 (2010) 1539.

**Figure captions**

**FIG. 1** (a) Schematics of β-Mn-type crystal structures with space group $P4_332$ and its enantiomer $P4_132$ through mirror symmetry. The actual space group depends on the chirality. Blue and red spheres represent $12d$ and $8c$ Wickoff sites, respectively[34]. (b) the Rietveld refinement of the XRD pattern for $Co_7Zn_8Mn_5$ alloy at room temperature.

**FIG. 2** (a) Temperature dependence of magnetization $M(T)$ curves of $Co_7Zn_8Mn_5$ alloy upon ZFC and FC runs under a magnetic field of 200 Oe. (b) Isothermal $dM/dH$ versus $\mu_0H$ curves at different temperatures (shifted up for clarity).

**FIG. 3** (a) The real part of ac-susceptibility as a function of applied field with ac field amplitude of 0.3 mT and frequency $f = 331$ Hz at various temperatures after ZFC process (shifted vertically for clarity). The dark gray shade represents the narrow $T$-$H$ region of Sk phase. Magnetic field dependence of $dM/dH$ (b), $\chi'$ (c) and $\chi''$ (d) at 216 K. The black dotted lines correspond to the critical fields.

**FIG. 4** The ac ME coefficient $\alpha_x$ (a) and $\alpha_y$ (b) as a function of magnetic field at different temperatures. Temperature dependence of ME coefficient $\alpha_x$ (c) and $\alpha_y$ (d) at different applied fields.

**FIG. 5** The Argand diagram of ac ME coefficient for $Co_7Zn_8Mn_5$/PMN-PT laminate at different temperatures with a frequency of 997 HZ.

**FIG. 6** Transport properties of $Co_7Zn_8Mn_5$ alloy. (a) Temperature dependence of longitudinal resistivity under $\mu_0H = 0$ T. Magnetic field dependence of magnetoresistance (b) and Hall resistivity (c) at various temperatures. The illustration of (c) shows the process of extracting the topological Hall resistivity at 10 K. The black line represents the measured Hall resistivity data, The blue line is the curve fitted in the high field region according to the formula $R_0H+$

$S_A \rho_{xx}^2 M$, and the red line is the extracted $\rho_{xy}^T$ curve. (d) Magnetic field dependence of $\rho_{xy}^T$ at different temperatures. The inset indicates temperature dependence of the maximum of $\rho_{xy}^T$ at different temperatures and the corresponding saturated magnetic field.

**FIG. 1**

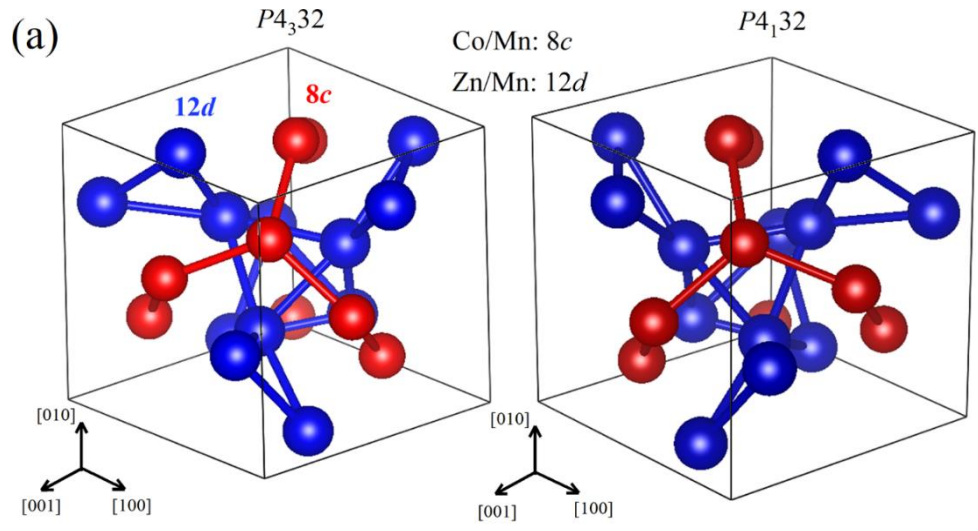

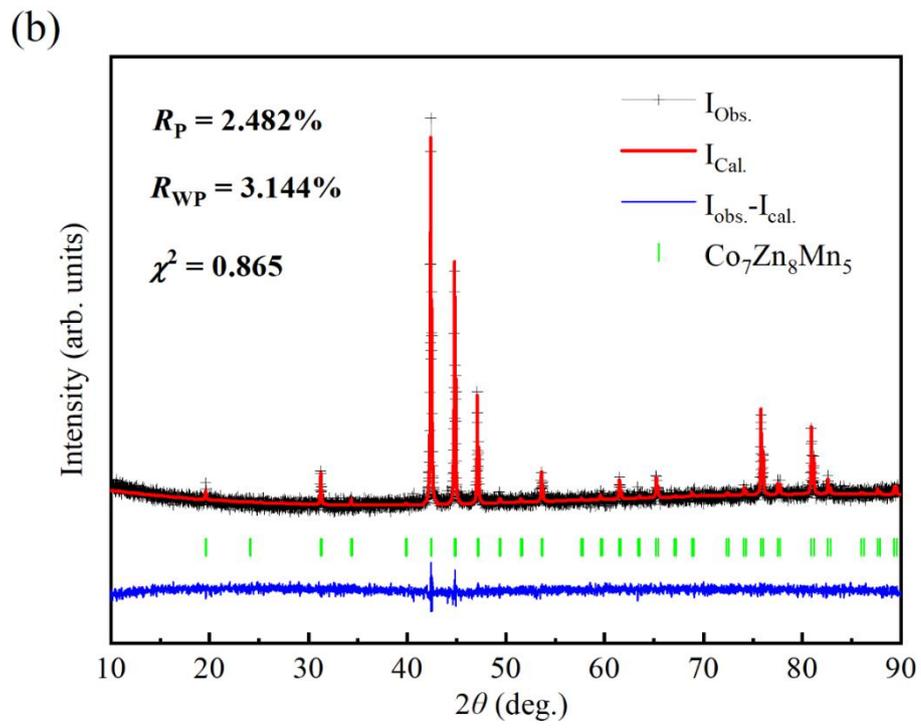

**FIG. 2**

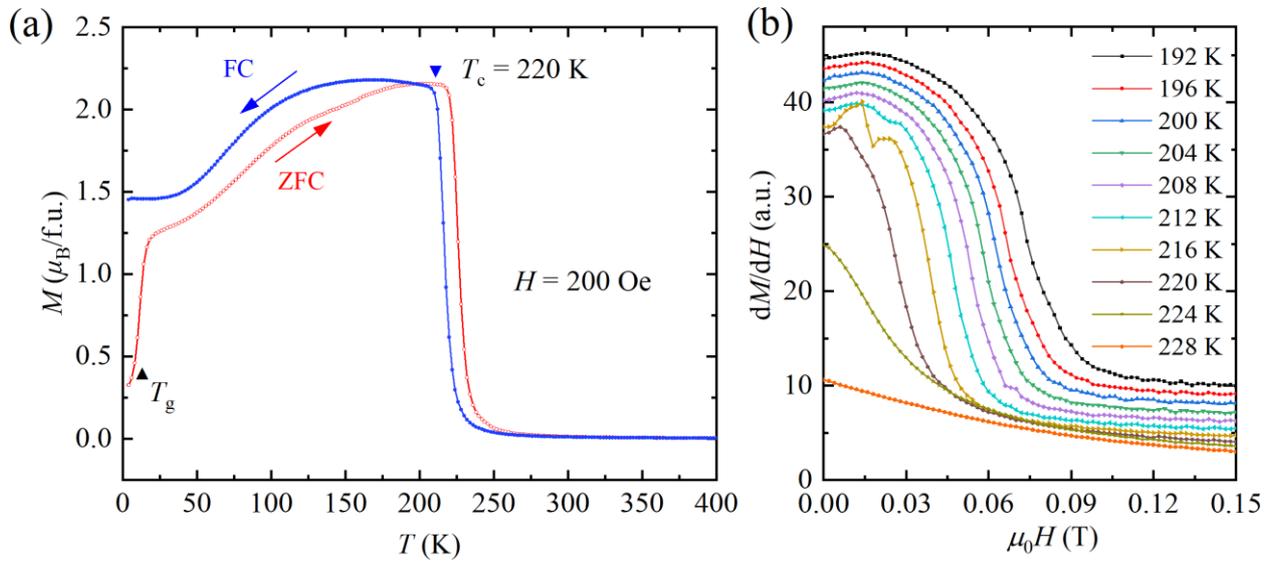

**FIG. 3**

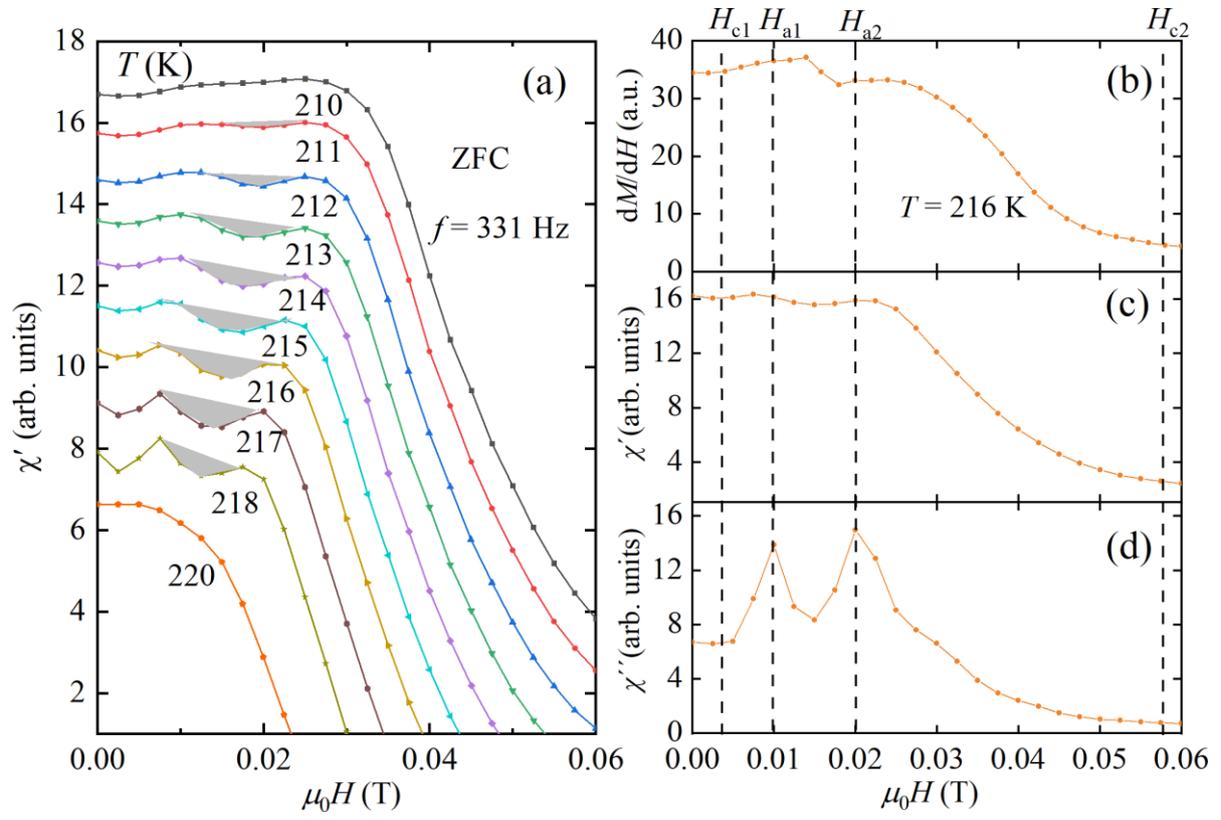

**FIG. 4**

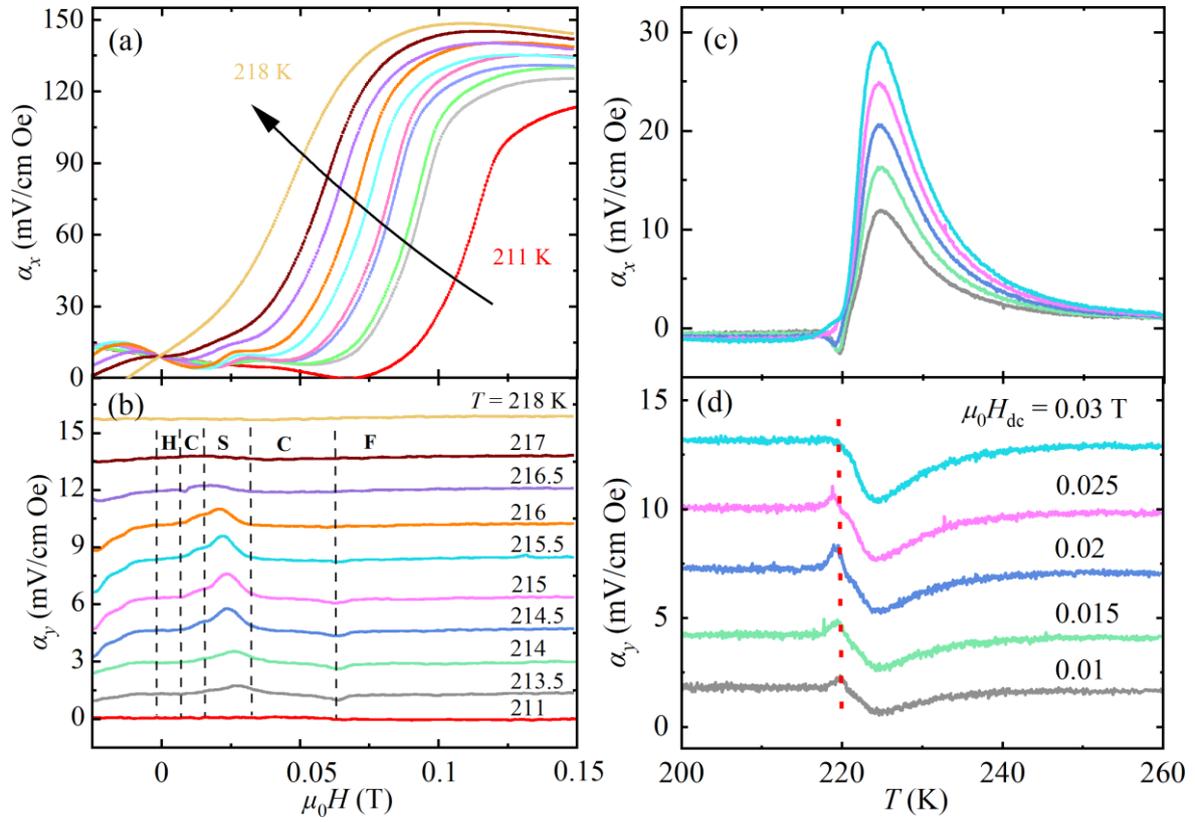

**FIG. 5**

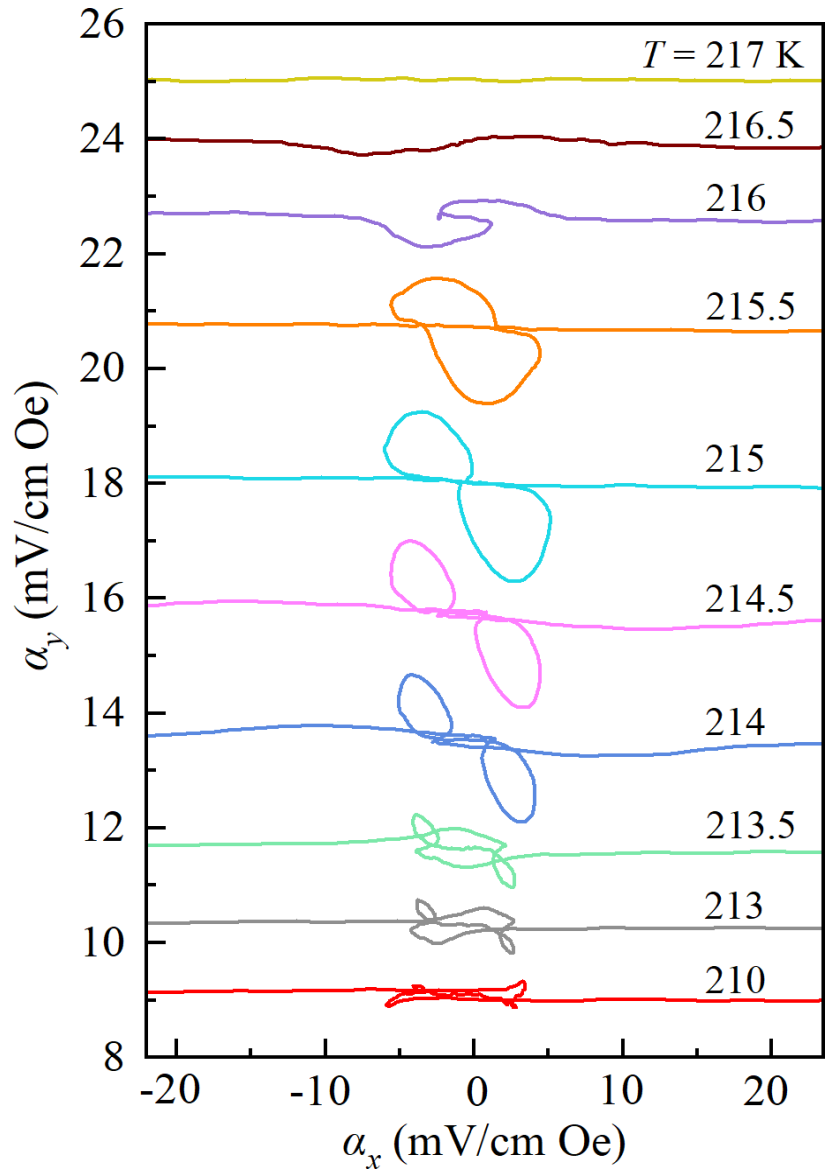

**FIG. 6**

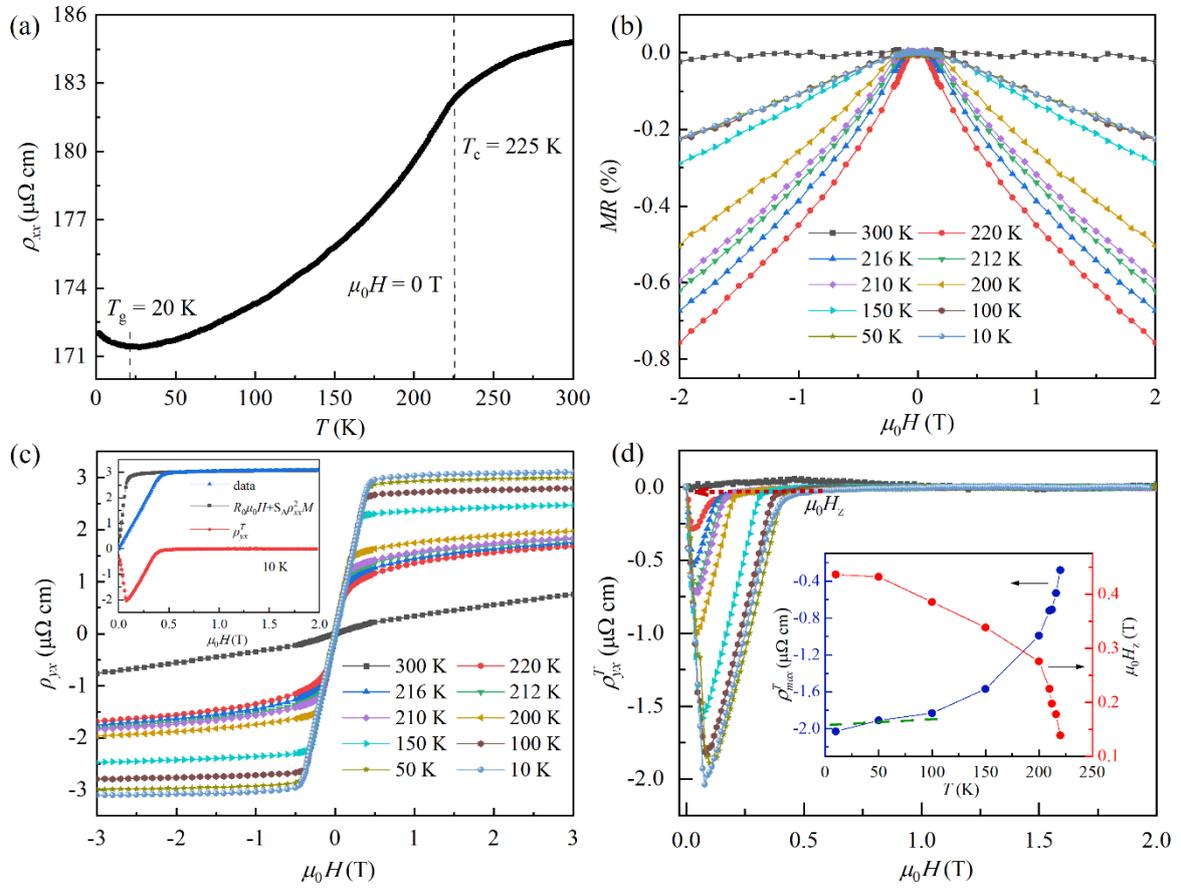

# Supplementary Material for

# Giant and robust topological Hall effect in Chiral Magnet Co$_7$Zn$_8$Mn$_5$

Hai Zeng[1,2,3,a], Xuanwei Zhao[1,2,3,4], Guang Yu[1,2,3,a], Xiaohua Luo[1,2], Shengcan Ma[1,2,3*], Changcai Chen[1,2], Zhaojun Mo[7], Yugang Zhang[5,6], Yisheng Chai[5,6,**], Jun Shen[7,8,9], and Zhenchen Zhong[1,2,3]

[1] Jiangxi Key Laboratory for Rare Earth Magnetic Materials and Devices, College of Rare earths, Jiangxi University of Science and Technology, Ganzhou 341000, People's Republic of China.

[2] Ganzhou Key Laboratory for Rare Earth Magnetic Functional Materials and Physics, College of Rare earths, Jiangxi University of Science and Technology, Ganzhou 341000, People's Republic of China.

[3] School of Materials Science and Engineering, Faculty of Materials Metallurgy and Chemistry, Jiangxi University of Science and Technology, Ganzhou 341000, People's Republic of China.

[4] China Key Laboratory of Radiation Physics and Technology and Key Laboratory of High Energy Density Physics and Technology of Ministry of Education, Sichuan University, Chengdu, 610064, China.

[5] Low Temperature Physics Laboratory, Department of Applied Physics, Chongqing University, Chongqing 401331, People's Republic of China.

[6] Beijing National Laboratory for Condensed Matter Physics, Institute of Physics, Chinese Academy of Sciences, Beijing 100190, People's Republic of China.

[7] Ganjiang Innovation Academy, Chinese Academy of Sciences, Ganzhou, Jiangxi 341000, People's Republic of China.

[8] Key Laboratory of Cryogenics, Technical Institute of Physics and Chemistry, Chinese Academy of Sciences, Beijing 100190, People's Republic of China.

[9] University of Chinese Academy of Sciences, Beijing 100049, People's Republic of China.

**Corresponding author:**

*Authors to whom correspondence should be addressed: shengcanma801@gmail.com, mashengcan@jxust.edu.cn (Shengcan Ma) and yschai@cqu.edu.cn (Yisheng Chai).

[a] These authors contributed equally: Hai Zeng, Guang Yu

## 1. Two-dimensional mappings from energy dispersive X-ray spectroscopy

The real chemical composition was determined to be $Co_{7.3}Zn_{8.0}Mn_{4.58}$ by using the energy dispersive x-ray spectroscopy (Fig. S1), which is very close to nominal starting composition of compound $Co_7Zn_8Mn_5$.

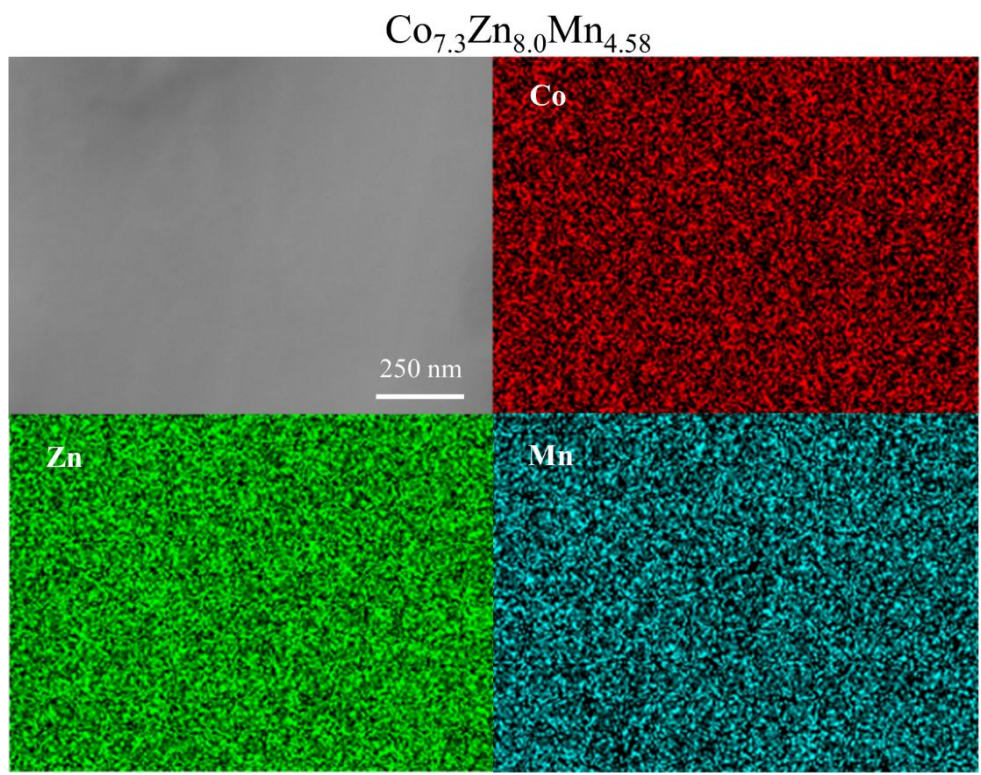

**FIG. S1** Two-dimensional mappings of constituent elements, Co, Zn, and Mn, in terms of Energy dispersive X-ray spectroscopy (EDX) for $Co_7Zn_8Mn_5$ alloy.

## 2. Isothermal hysteresis loop curves

The isothermal magnetization loops are measured and presented in Fig. S2. It is found that the magnetization reaches saturation very rapidly below $T_c$ with low saturated field, signifying the apparent ferromagnetic behavior. Furthermore, it is of significance that the magnetization behavior without a hysteresis is observed in the whole temperature range covered in this work.

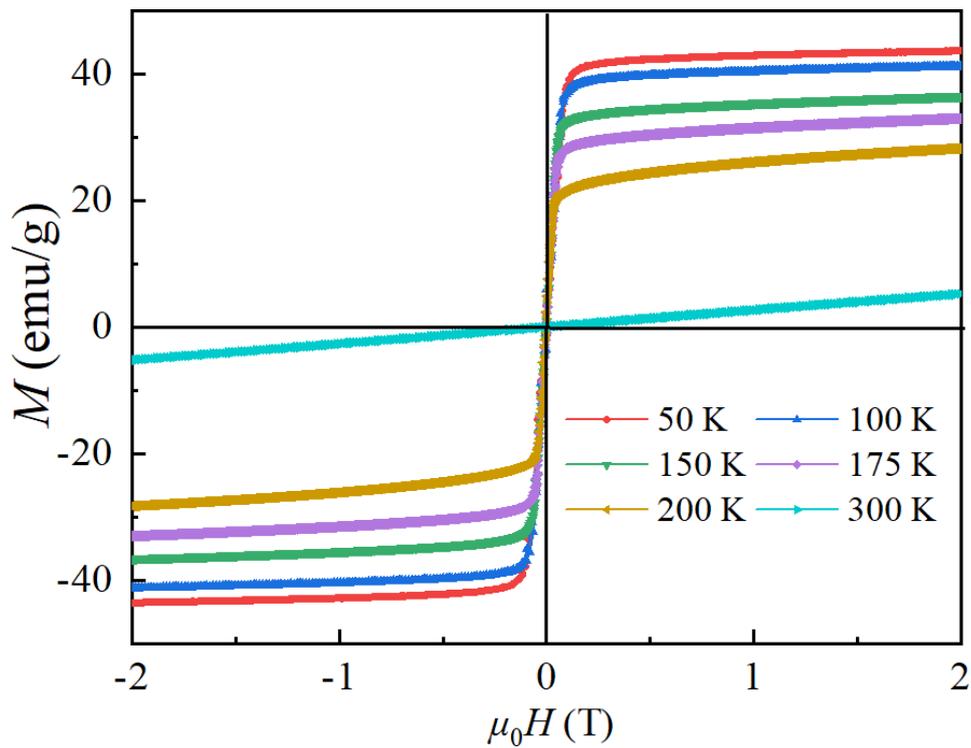

**FIG. S2** Isothermal hysteresis loop curves at different temperatures.

### 3. Ac susceptibility

Here, the evolution of the magnetic state with the magnetic field for the sample can be distinguished clearly. The results showed that the magnetic skyrmions survive in temperature interval of 210-220 K.

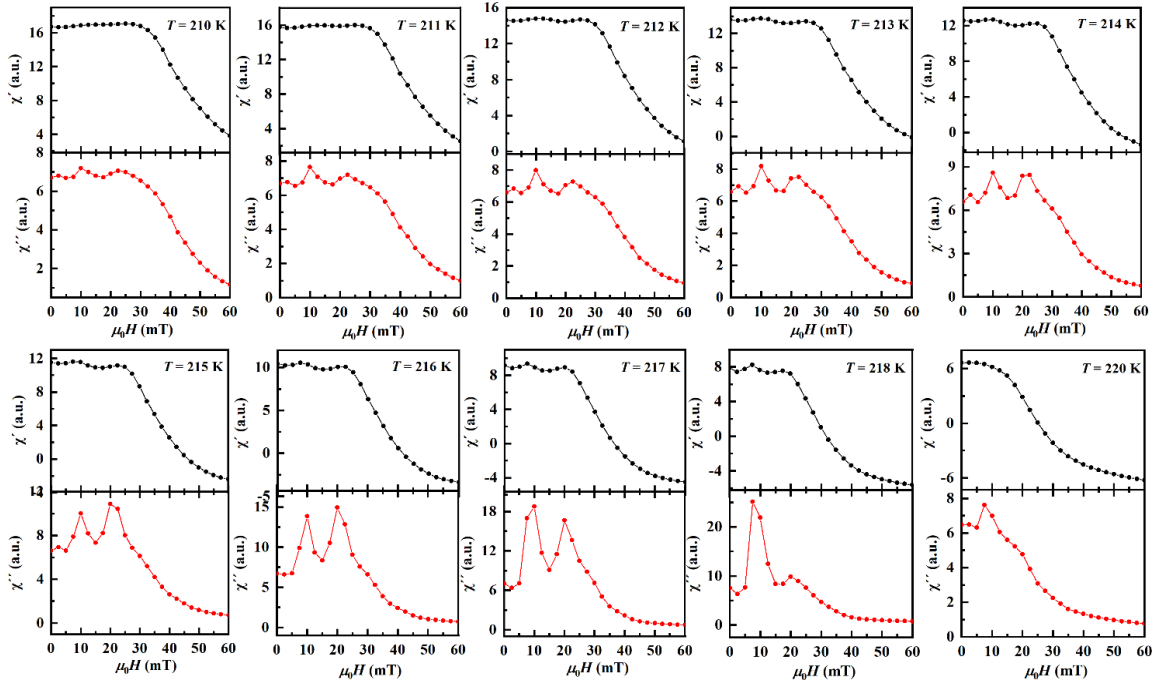

**FIG. S3** Magnetic-field dependence of $\chi'$ and $\chi''$ of $Co_7Zn_8Mn_5$ alloy for some selected temperatures after zero-field cooling.